\documentclass[12pt,preprint]{aastex}
\usepackage{psfig,epsfig,amsfonts,amsmath,amssymb,graphicx,morefloats,appendix,tensor}
\usepackage{color}
\usepackage{natbib}
\usepackage{hyperref}
\begin{document}

\slugcomment{Accepted to ApJL: May 15, 2019}

\title{Multiple Rings of Millimeter Dust Emission in the HD 15115 Debris Disk
}

\author{Meredith A. MacGregor\altaffilmark{1,2}, Alycia J. Weinberger\altaffilmark{1}, Erika R. Nesvold\altaffilmark{1}, A. Meredith Hughes\altaffilmark{3}, D. J. Wilner\altaffilmark{4}, Thayne Currie\altaffilmark{5}, John H. Debes\altaffilmark{6}, Jessica K. Donaldson\altaffilmark{1}, Seth Redfield\altaffilmark{3}, Aki Roberge\altaffilmark{7}, Glenn Schneider\altaffilmark{8}}

\altaffiltext{1}{Department of Terrestrial Magnetism, Carnegie Institution for Science, 5241 Broad Branch Road NW, Washington, DC 20015, USA}
\altaffiltext{2}{NSF Astronomy and Astrophysics Postdoctoral Fellow}
\altaffiltext{3}{Astronomy Department and Van Vleck Observatory, Wesleyan University, 96 Foss Hill Drive, Middletown, CT 06459, USA}
\altaffiltext{4}{Harvard-Smithsonian Center for Astrophysics, 60 Garden St., Cambridge, MA 02138, USA}
\altaffiltext{5}{National Astronomical Observatory of Japan, Subaru Telescope, National Institutes of Natural Sciences, Hilo, HI 96720, USA}
\altaffiltext{6}{Space Telescope Science Institute, 3700 San Martin Drive, Baltimore, MD, 21218, USA}
\altaffiltext{7}{Exoplanets and Stellar Astrophysics Lab, NASA Goddard Space Flight Center, Greenbelt, MD 20771, USA}
\altaffiltext{8}{Steward Observatory, The University of Arizona, 933 North Cherry Avenue, Tucson, AZ 85721, USA}

\begin{abstract}

We present observations of the HD~15115 debris disk from ALMA at 1.3 mm that capture this intriguing system with the highest resolution ($0\farcs6$ or $29$~AU) at millimeter wavelengths to date.  This new ALMA image shows evidence for two rings in the disk separated by a cleared gap.  By fitting models directly to the observed visibilities within a MCMC framework, we are able to characterize the millimeter continuum emission and place robust constraints on the disk structure and geometry.  In the best-fit model of a power law disk with a Gaussian gap, the disk inner and outer edges are at $43.9\pm5.8$~AU ($0\farcs89\pm0\farcs12$) and $92.2\pm2.4$~AU ($1\farcs88\pm0\farcs49$), respectively, with a gap located at $58.9\pm4.5$~AU ($1\farcs2\pm0\farcs10$) with a fractional depth of $0.88\pm0.10$ and a width of $13.8\pm5.6$~AU ($0\farcs28\pm0\farcs11$).  Since we do not see any evidence at millimeter wavelengths for the dramatic east-west asymmetry seen in scattered light, we conclude that this feature most likely results from a mechanism that only affects small grains.  Using dynamical modeling and our constraints on the gap properties, we are able to estimate a mass for the possible planet sculpting the gap to be $0.16\pm0.06$~$M_\text{Jup}$.

\end{abstract}


\section{Introduction}
\label{sec:intro}

As planets form in circumstellar disks, they inherit their dynamics and composition, and imprint their presence on the remaining material through dynamical interactions.  Debris disks, the end-stage of circumstellar evolution, are continually replenished through collisions between remnant asteroids and comets that produce dust grains over a broad range of sizes. Small micron-sized grains are influenced by non-gravitational effects including radiation pressure and interactions with the interstellar medium (ISM).  Larger millimeter-sized grains are less affected by these forces, making them more reliable tracers of the underlying planetesimal belt structure.  Both scattering albedo and thermal emission are strong functions of grain size; grains emit at wavelengths comparable to their sizes.  Thus, millimeter wavelength observations probe larger grains and provide the best opportunity to detect the influence of planets on surrounding disk material.  

HD~15115 hosts a debris disk initially detected as a strong infrared excess \citep{Silverstone:2000,Moor:2006} and later revealed in scattered light with the \emph{Hubble Space Telescope} (HST) to have an edge-on disk with pronounced asymmetries \citep{Kalas:2007,Debes:2008,Schneider:2014}. Since then, the disk has been imaged multiple times in scattered light with higher resolution by Keck, LBT, Gemini, Subaru, and VLT \citep{Rodigas:2012,Mazoyer:2014,Sai:2015,Engler:2018}.   In most scattered light images, the west side of the disk extends nearly twice as far from the star and has a higher flux.  The star is mid-F type \cite[estimates range from F2-F4,][]{Ochsenbein:1980,Harlan:1974} at a GAIA DR2 distance of $49.0\pm0.1$~pc ($\sim10\%$ farther than the Hipparcos distance of 45~pc). The age of the system is not well-known, and is derived from potential membership in young moving groups including the $\sim20$ Myr-old $\beta$ Pictoris Association \citep{Moor:2006} and the $\sim45$~Myr-old Tucana-Horologium Association \cite[from \texttt{BANYAN} $\Sigma$ using GAIA RV,][]{Gagne:2018}.

Previous millimeter observations of the HD~15115 debris disk hinted that the asymmetric structure seen in scattered light might also be traced by larger grains, but lacked the sensitivity or resolution to draw firm conclusions \citep{MacGregor:2015a}.  Here, we present new ALMA observations of this intriguing system with the highest resolution and sensitivity at these wavelengths to date that trace the large grain population, and by inference, the planetesimal locations within the disk.

\section{Observations}
\label{sec:obs}

We observed the HD~15115 debris disk with ALMA in Cycle 3 using Band 6 (1.3~mm, 230~GHz).  Two scheduling blocks (SBs) were executed in both a compact (baselines of 15--310~m) and more extended (baselines of 15--704~m) configuration with 36 antennas in the array on 2016 January 1 and 2016 June 9, respectively.  The total observing duration was 31.7~min with an on-source time of 15.1~min, and 48.5~min with 30.2~min on-source for the compact and extended configurations, respectively.  For both executions, the precipitable water vapor (PWV) was $<1.5$~mm.  

The correlator was set-up to maximize continuum sensitivity while still covering the $^{12}$CO J$=2-1$ line at 230.538~GHz with high spectral resolution.  To achieve this, we used four spectral windows with central frequencies of 230.538 (centered on the $^{12}$CO J$=2-1$ line), 232.538, 215.5, and 217.5~GHz.  The three continuum-only spectral windows had a bandwidth of 2~GHz with 128 channels, while the final spectral window had a reduced bandwidth of 1.875~GHz with 3840 channels.  

Both the compact and more extended SBs made use of the same calibration sources.  The bright blazar J0238+1636 ($10\fdg8$ away from the target) was used for both bandpass and flux calibration, as well as pointing.  We estimate that the absolute flux calibration uncertainty is $<10\%$.  Observations of J0224+0659 ($0\fdg8$ away) were interleaved with the target to account for time-dependent gain variations due to instrumental and atmospheric effects.  All data processing and calibration was done with the ALMA pipeline in \texttt{CASA} (version 4.7.2).  To reduce data volume, the calibrated visibilities were time-averaged in 30~sec intervals.  All images were generated using the \texttt{CLEAN} task in \texttt{CASA}.

\section{Results and Analysis}
\label{sec:results}

\begin{figure}[t]
\begin{minipage}[h]{0.52\textwidth}
  \begin{center}
       \includegraphics[scale=0.8]{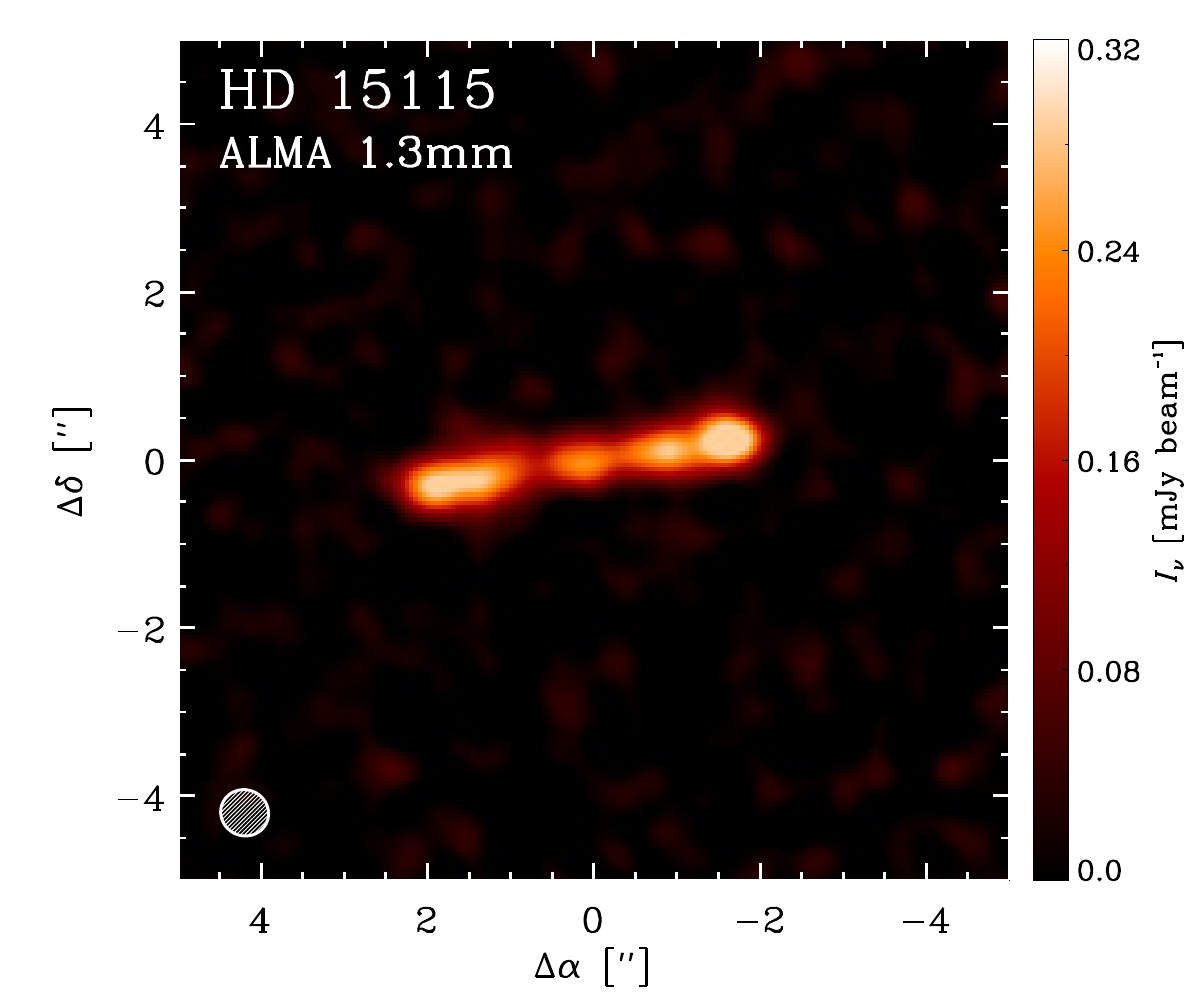}
  \end{center}
 \end{minipage}
\begin{minipage}[h]{0.48\textwidth}
  \begin{center}
       \includegraphics[scale=0.8]{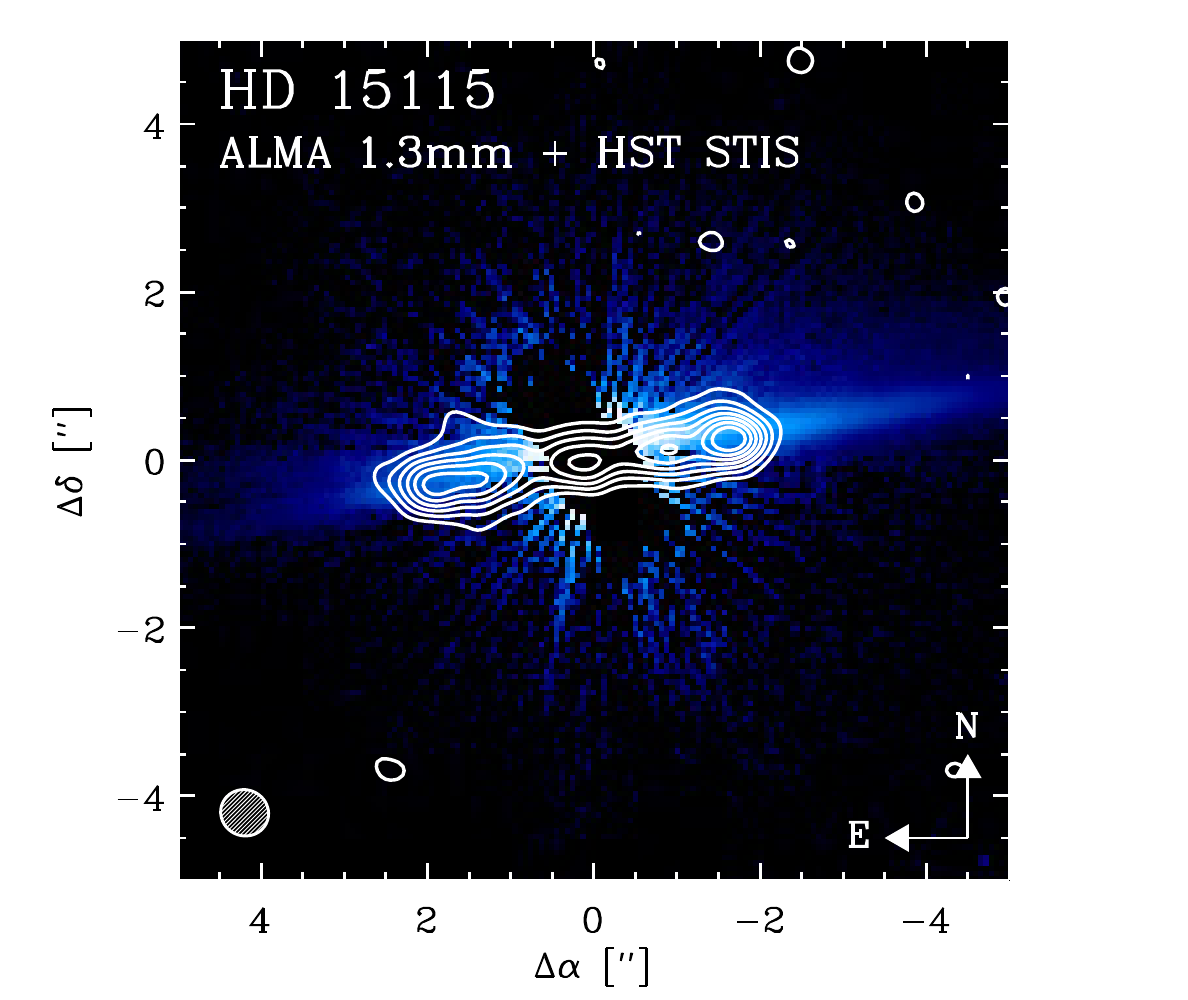}
  \end{center}
 \end{minipage}
 
\caption{\small  Our new ALMA 1.3~mm continuum image of HD~15115 (left)  shows evidence for two rings in the disk.  At right, the ALMA image is overlaid as $3\sigma$ contours ($3\times$ the rms noise of 15~$\mu$Jy~beam$^{-1}$) on the HST STIS image from \cite{Schneider:2014}.  In both panels, the white ellipse in the lower left corner indicates the synthesized beam size of $0\farcs58\times0\farcs55$ (with robust $=0.5$ weighting).
}
\label{fig:fig1}
\end{figure}

Our new ALMA 1.3~mm dust continuum image of HD~15115 is shown in Figure~\ref{fig:fig1} (left panel).  The synthesized beam is $0\farcs58\times0\farcs55$ ($28\times27$~AU) with robust $=0.5$ weighting, and the rms noise in the image is $15$~$\mu$Jy~beam$^{-1}$.  The right panel of Figure~\ref{fig:fig1} shows the same continuum image overlaid as contours in increments of $3\sigma$ over the previous HST STIS scattered light image.  With ALMA, we detect the disk and a point source coincident with the stellar position at $18\sigma$ and $15\sigma$, respectively.  Most surprisingly, the disk appears to consist of two rings separated by a gap at $\sim1\arcsec$ seen as two peaks or ansae (the characteristic limb brightening of an optically thin, edge-on disk) on either side of the star.  Overall, the millimeter emission aligns well with the scattered light, but does not exhibit the same radial extent.  Although the western side of the disk appears $\sim3\sigma$ brighter than the eastern side, there is no significant evidence for the dramatic east-west asymmetry seen in scattered light images.

We do not detect any $^{12}$CO emission from the disk in our observations.  Given this non-detection, we can determine a $3\sigma$ upper limit on the integrated flux density of the line of $<0.022$~Jy~km~s$^{-1}$ or $<1.6\times10^{-22}$~W~m$^2$, which implies an upper limit on the total CO mass of $<1.4\times10^{-9}$~$M_\oplus$ assuming Local Thermodynamic Equilibrium (LTE).  For comparison, this result is significantly below the upper limit of $<7.7\times10^{-7}$~$M_\oplus$ determined for HD~61005 \citep{Olofsson:2016} and the detected CO mass of $4.34\times10^{-4}$~$M_\oplus$ for HD~32297 \citep{MacGregor:2018b}, both assuming LTE.

\subsection{Modeling Approach}
\label{sec:model_approach}

\begin{figure}[t]
  \begin{center}
       \includegraphics[scale=0.4]{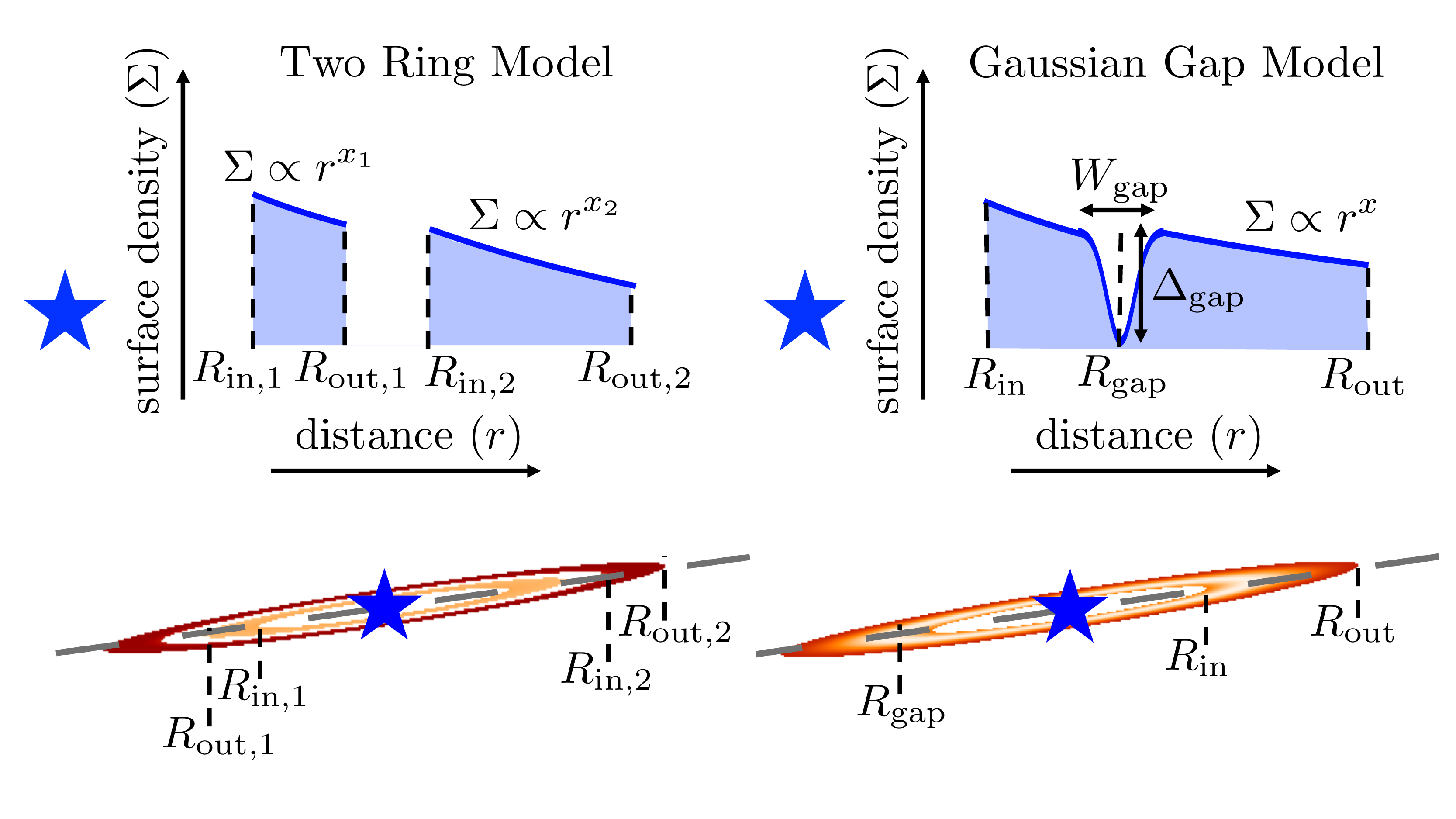}
  \end{center}
\caption{\small A schematic of both the two ring and Gaussian gap models.  The top images show the radial surface density profile for each disk model, while the bottom images show the resulting models at full resolution.
}
\label{fig:fig2}
\end{figure}

We fit three different models to the HD~15115 ALMA data-- a single ring model and two different double ring models.  The single ring model fits the outer regions of the disk well, but leaves significant residuals  at the locations of the inner ansae.  The double ring models consist of either (1) two power law rings with a completely empty gap between them, or (2) a single power law ring with a partially depleted, Gaussian gap.  The most significant difference between these two models is edge sharpness and fractional depth of the gap.  In all models, we assume that the surface brightness is an axisymmetric function, $I_\nu\propto r^\alpha$, where the power law index $\alpha$ incorporates both a temperature fall off, $T\propto r^{-0.5}$, and a surface density distribution, $\Sigma \propto r^{x}$, with two power law indices for the two ring model ($x_1$ and $x_2$) and one index for the gap model ($x$). The Gaussian gap is defined by a multiplicative function

\begin{equation}
    G_\text{gap}(r) = 1- \Delta_\text{gap}\times\text{exp}\left[\frac{-(r-R_\text{gap})^2}{2\sigma_\text{gap}^2}\right],
\end{equation}

\noindent where $\Delta_\text{gap}$ is the fractional depth, $R_\text{gap}$ is the radial location, and the FWHM is $W_\text{gap}=2\sqrt{2\text{ln}2}\sigma_\text{gap}$, all of which are free model parameters.  Figure~\ref{fig:fig2} shows the surface density profiles for both models along with full resolution images.  The total flux density of the disk is normalized to $F_\text{disk}=\int I_\nu d\Omega$, and the flux of the central point source is $F_\text{pt}$.  We also fit for the disk geometry, namely the inclination ($i$) and position angle ($PA$), as well as an offset in RA and DEC of the star from the disk centroid ($\Delta\alpha$ and $\Delta\delta$).

In order to efficiently explore the parameter space and characterize the uncertainties, we adopt the modeling procedure initially described in \cite{MacGregor:2013} and most recently in \cite{MacGregor:2018b}, where models are fit to the millimeter visibilities within a Markov Chain Monte Carlo (MCMC) framework.  We make use of both the \texttt{emcee} \citep{Foreman-Mackey:2013} and \texttt{vis\_sample}\footnote{\texttt{vis\_sample} is publicly available at \url{https://github.com/AstroChem/vis_sample} or in the Anaconda Cloud at
\url{https://anaconda.org/rloomis/vis_sample.}} python packages.  We assume uniform priors for all parameters and only introduce limits to ensure that each model is physical: $F_\text{disk}>0$ and $0<R_\text{in}<R_\text{out}$.  We use $\sim10^6$ iterations (100 walkers, 10,000 steps each) to fully explore parameter space, and evaluate the fit quality of each model using a $\chi^2$ likelihood function, $\text{ln}\mathcal{L}=-\chi^2/2$.  To check for convergence, we examine all chains and compute the Gelman-Rubin statistic \citep{Gelman:1992} requiring $\hat{R}<1.1$ for all model parameters. The one-dimensional marginalized probability distributions for all model parameters appear Gaussian.  We do note a degeneracy between the surface density gradient and inner/outer disk radii as has been discussed in previous works \cite[e.g.,][]{MacGregor:2015a}.  The $1\sigma$ errors are determined by assuming normally distributed errors where the probability that a measurement has a distance less than $a$ from the mean value is given by $\text{erf}\left(\frac{a}{\sigma\sqrt2}\right)$.

\subsection{Modeling Results}
\label{sec:model_results}

\begin{figure}[t]
  \begin{center}
       \includegraphics[scale=0.95]{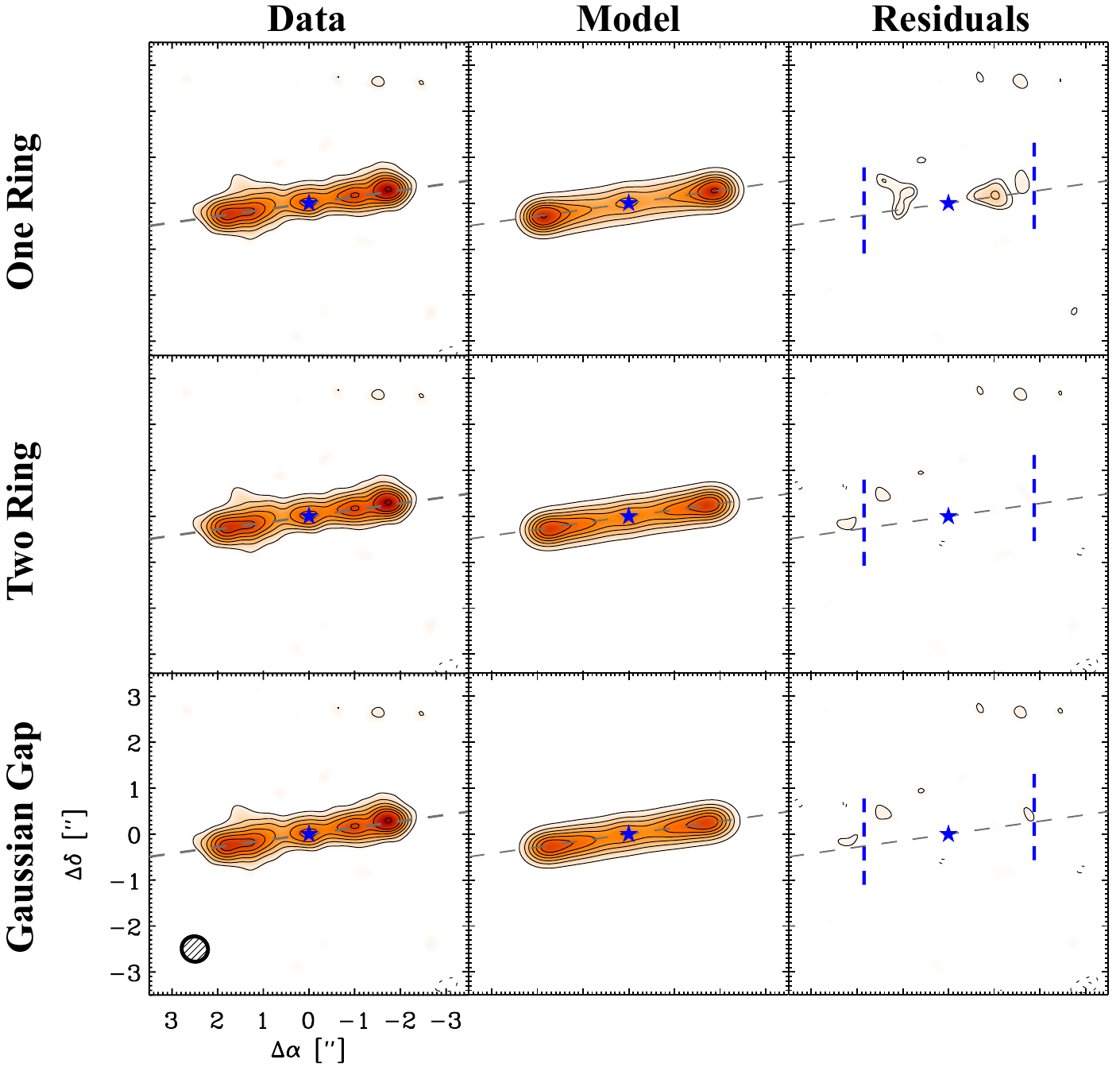}
  \end{center}
\caption{\small The best-fit two ring (middle) and Gaussian belt models (bottom) both agree well with the data.  A model with only one ring (top) leaves significant residuals indicating the need for an additional model component.  For all rows, the ALMA 1.3~mm continuum image is shown at left, the best-fit model imaged like the data is shown in the middle, and the resulting residuals imaged like the data are shown at right.  In all panels, contours are in steps of $3\sigma$ contours ($3\times$ the rms noise of 15~$\mu$Jy~beam$^{-1}$), except in the residual images where a $2\sigma$ contour has been added. The location of the central star is indicated by the blue star symbol, and the dashed blue vertical lines in the residual panels mark the radial position of the outer ring.  The ellipse in the lower left panel shows the synthesized beam size of $0\farcs58\times0\farcs55$ (same as in Figure~\ref{fig:fig1} with robust $=0.5$ weighting).
}
\label{fig:fig3}
\end{figure}

Both the two ring and gap models provide good fits to the data with reduced $\chi^2$ values of $\sim1.1$.  Table~\ref{tab:results} lists the best-fit parameter values for both models, and Figure~\ref{fig:fig3} shows the data, the best-fit models imaged like the data, and the resulting residuals.  The residuals are minimal ($\leq3\sigma$) for both models.  For comparison, the first row of Figure~\ref{fig:fig3} shows the results for the single ring model, which leaves significant ($6\sigma$) residuals close to the star, indicating the need for an additional model component.  Given the current resolution of these observations, we cannot distinguish between the two ring and Gaussian gap models.  However, the best-fit Gaussian gap model has a fractional depth of $0.88\pm0.10$, implying an almost completely empty gap.  Since both models yield this result, we conclude that the gap between the inner and outer rings must be nearly devoid of millimeter emission.

Despite using different parameterizations, both best-fit models yield consistent radial locations for the two rings and gap.  In the two ring model, the inner ring is located between $44.6\pm4.5$~AU and $50.9\pm8.8$~AU, and the outer ring spans from $65.7\pm4.5$~AU to $92.8\pm3.1$~AU, implying a gap at $58.3\pm2.2$~AU with a width of $14.8\pm4.3$~AU.  In the Gaussian gap model, the inner edge of the disk is at $43.9\pm5.8$~AU and the outer edge is at $92.8\pm3.2$.  The gap is located at $58.9\pm4.5$~AU with a FWHM of $13.8\pm5.6$~AU, nearly identical to the results from the two ring model.  The best-fit total disk and stellar flux densities, the offset of the star from the disk centroid, and the disk geometry (inclination and position angle) are also consistent between both models.  For the two ring model, the flux densities of the inner and outer rings are $0.38\pm0.08$~mJy and $1.61\pm0.09$~mJy, respectively.  The combined flux is $1.99\pm0.10$~mJy, which is nearly identical to the total flux determined from the Gaussian gap model of $1.98\pm0.03$~mJy.  Neither model places a strong constraint on the surface density profile of the disk or the gap edge sharpness.  The best-fit value for the flux of the central star is $0.04\pm0.01$~mJy, about $2\sigma$ in excess of the expected flux of the stellar photosphere.  We attribute this slight difference to chromospheric emission as has been seen for stars with similar spectral types including $\alpha$ Centauri A/B and $\epsilon$ Eridani \citep{Liseau:2016,MacGregor:2015b}.  There is no evidence for a significant offset between the star and the disk centroid, indicating that any eccentricity of the disk must be small.

\section{Discussion}
\label{sec:disc}

We have presented a new high-resolution image of the HD~15115 debris disk from ALMA at 1.3~mm that shows evidence for multiple ring structure.  Here, we compare this ALMA image to previous studies of the same system, discuss disk sculpting mechanisms, and place constraints on a possible planet opening the observed gap.

\subsection{Comparison to Previous Observations}
\label{sec:comparison}

 In previous scattered light images, the HD~15115 debris disk shows an extreme asymmetry, with the eastern side of the disk extending to only $\sim7\arcsec$ and the western side reaching $>12\arcsec$ \citep{Kalas:2007,Debes:2008,Schneider:2014,Mazoyer:2014,Sai:2015,Engler:2018}, although \cite{Rodigas:2012} do not see an asymmetry at 3.8~$\mu$m.  The disk also appears bowed, which is thought to result from an inclined ($86-87\degr$), highly forward scattering disk.  An additional halo of small dust grains (1) extends up to 620~AU from the star on its western side, (2) appears to slope away from the disk major axis on the eastern side, and (3) appears out (north) of the disk plane on the western side in HST images \citep{Schneider:2014}.  We do not see any of these features in our ALMA image, implying that they are only traced by the small, micron-sized grains that dominate scattered light images.  We have modified the results of previous work to bring them all to the same scale, at the distance provided by the new GAIA parallax ($49.0\pm0.1$~pc).

Our best-fit values for the disk inner and outer edges agree with previous determinations.  \cite{Debes:2008} and \cite{Rodigas:2012} detect steepening of the surface brightness profile for the western side of the disk between $\sim87-98$~AU, consistent with our determination of $\sim92$~AU for the outer disk edge.  \cite{Mazoyer:2014} show conclusively that the outer ring is symmetrical with a radius of $\sim2\arcsec$ ($\sim98$~AU).  \cite{Schneider:2014} and \cite{Mazoyer:2014} note a partial clearing of the disk interior to $1\arcsec$ (49~AU), and \cite{Moor:2011} determine a minimum radius of $46\pm2$~AU from SED modeling.  Our best-fit inner radius is somewhat interior to these values at $\sim44$~AU, but consistent within the uncertainties.  Low-resolution millimeter imaging with the Submillimeter Array (SMA) gave comparable values of $120^{+31}_{-22}$~AU and $47\pm28$~AU for the inner and outer radius, respectively \citep{MacGregor:2015a}.

These new ALMA observations are the first to resolve and conclusively determine the presence of a second dust ring in the HD~15115 debris disk.  However, several previous studies have suggested that this system might contain multiple rings of dust.  The SED fits of \cite{Moor:2011} included a hot dust component at $4\pm2$~AU.  Recently, \cite{Engler:2018} found evidence in SPHERE images for a second ring interior to $\sim1\farcs3$ ($\sim63$~AU), slightly outside of our best-fit gap position of $\sim59$~AU.  Our ALMA observations do not indicate that the inner ring is misaligned with the outer ring as \cite{Engler:2018} suggest.  While the disk is not vertically resolved in our image, a misalignment of $\sim6\degr$ should have been detectable.

\subsection{Possible Sculpting Mechanisms}
\label{sec:sculpting}

The origin of HD~15115's complex scattered light structure has been much debated.  There are similarities to other edge-on, asymmetric systems such as HD~61005 and HD~32297, whose structure was originally attributed to ISM interactions \citep{Schneider:2014} but which recent ALMA observations show might have a planetary origin \citep{MacGregor:2018b}.  Although HD~15115 has a stellar luminosity intermediate to HD 32297 and HD 61005, it has a distinctly different structure pointing to a different dynamical origin.  By considering the resolved structure of the HD~15115 debris disk in both scattered light (small grains) and millimeter (large grains) images, we can place new constraints on how this system is shaped.  Notably, since the ALMA image does not show any asymmetry $>3\sigma$ between the eastern and western sides of the disk, we conclude that the mechanism producing this asymmetry likely only operates on small grains.  Many possible mechanisms have been suggested previously including stellar encounters \citep{Kalas:2007}, local increases in collisions \citep{Mazoyer:2014}, and interactions with the interstellar medium \cite[ISM,][]{Debes:2009}, which we consider here. 

Stellar encounters could affect the orbits of planetesimals in a disk and generate complex structures as they continue to evolve and collide. \cite{Kalas:2007} suggested that interaction with the nearby M star HIP~12545, which differs from HD 15115 by only 3.5 km/s in UVW and is currently only 5.5~pc away, could have shaped the disk $\sim1.5$~Myr ago.  However, \texttt{BANYAN} $\Sigma$ does not find a high probability that HIP 12545 is in Tuc-Hor.  The stars are currently moving apart in X and closer together in Y and Z, making a previous encounter likely only if the stars were bound, which is itself unlikely given their large separation.  We note that eccentric planets within a disk can have a similar effect on disk structure as a stellar encounter, and modeling has shown that high eccentricity planets can generate asymmetric disks \cite[e.g.,][]{Lee:2016}.
  
Perhaps the most likely explanation for the scattered light asymmetry is an interaction with the ISM, which operates predominantly on small grains and is unlikely to affect large grains.  In this scenario, the eastern side of the disk becomes truncated when it impacts a dense clump of interstellar gas.  Ram pressure from the interaction strips grains from the disk into an extended halo on the western side of the disk.  Conveniently, the proper motion of HD~15115 is almost entirely along the major axis of the disk (towards the east).  No absorption is detected along the line of sight to HD~15115 in CaII \citep{Iglesias:2018}, although more sensitive tracers may reveal weak circumstellar and/or interstellar absorption.  Disk gas could also be stripped by ISM interactions and then remove grains when they become entrained \citep{Maness:2009}.  However, we do not detect any $^{12}$CO emission in our ALMA observations, which indicates that there is likely insufficient disk gas for this process to occur.

\subsection{Constraints on a Planet in the Gap}
\label{sec:planet}

\begin{figure}[t]
  \begin{center}
       \includegraphics[scale=0.65]{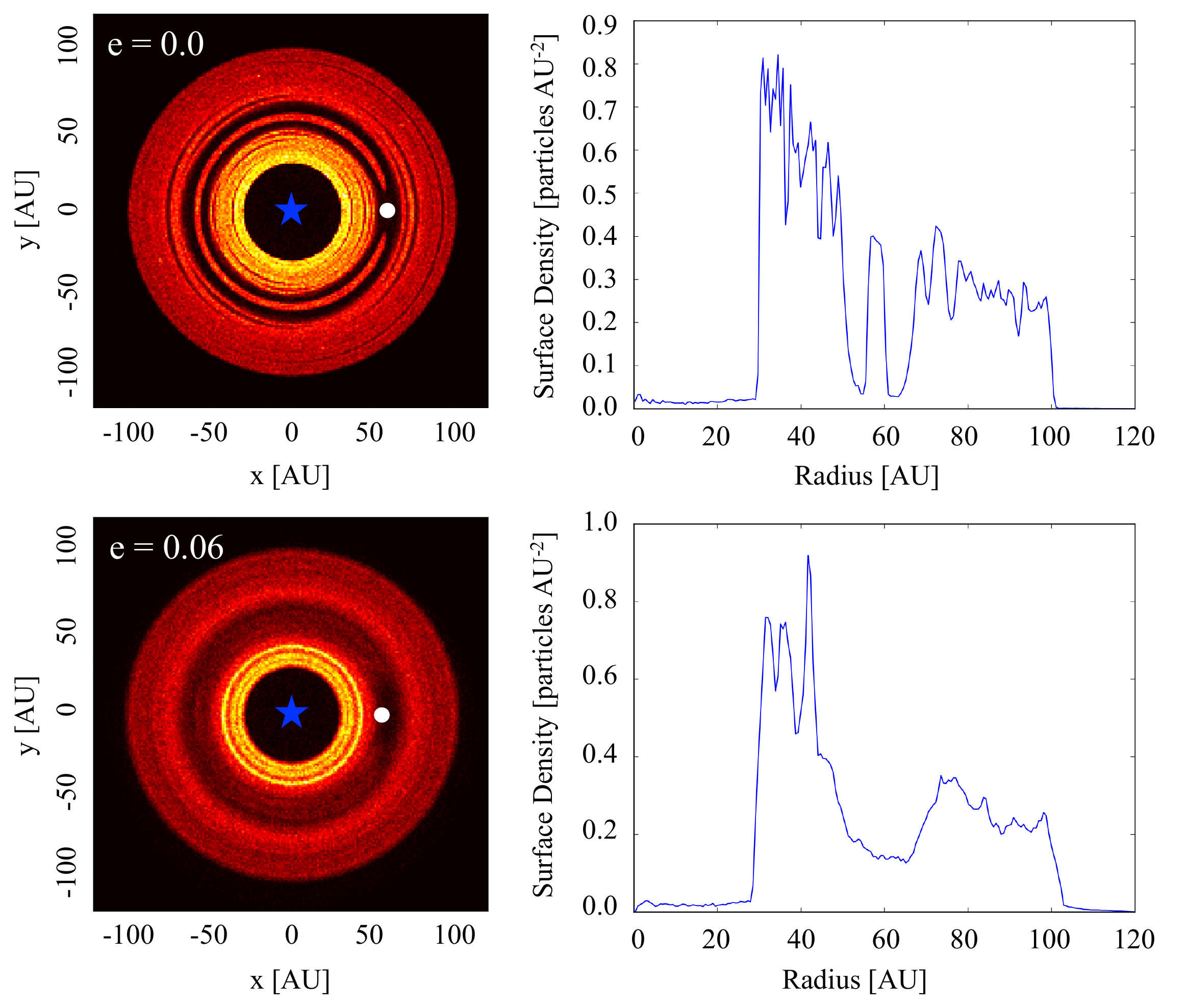}
  \end{center}
\caption{\small REBOUND simulations show that a $0.2$~$M_\text{Jup}$-mass planet with semi-major axis 58~AU could open the observed gap in the HD~15115 debris disk within 1~Myr.  The face-on disk image \emph{(left)} and radial surface density profile \emph{(right)} are shown for planets with eccentricity $e=0$ \emph{(top)} and $e=0.06$ \emph{(bottom)}.  Both models assume a 1.4~$M_\odot$ star and include 10000 particles.  In the face-on images, the stellar position is marked by the blue star and the planet pericenter position with the white circle at 58~AU and 54.52~AU in the circular and eccentric simulations, respectively.
}
\label{fig:fig4}
\end{figure} 

The ALMA image of HD~15115 shows two bright ansae on both sides of the central star, and our modeling favors the interpretation that this structure results from two rings of dust separated by a depleted gap.  Before concluding that this surface brightness distribution stems from a planetary origin, we must consider other possibilities.  If particles in the disk were on eccentric orbits but not apsidally aligned, models predict that particles would pile up on the disk inner and outer edges making them appear brighter \cite[i.e.,][]{Pan:2016}.  For a nearly edge-on disk, the resulting surface brightness distribution would show two bright peaks on either side of the star, comparable `by eye' to our ALMA image.  To test this, we created toy disk models using particles with fixed semi-major axis and eccentricity, but random longitude of periapses.  While these models do produce the predicted effect, the maximum brightness difference between the two apparent peaks is $\sim40-50\%$, not large enough to match our observations.   

We conclude that the most likely mechanism to produce the observed structure of the HD~15115 debris disk is removal of planetesimals via a planet in the gap.  The Gaussian gap model constrains the gap position and width to be $58.9\pm4.5$~AU and $13.8\pm5.6$~AU, respectively, with a fractional depth of $0.88\pm0.10$. We can use these best-fit parameter values to place constraints on the mass of the potential planet.  \cite{Quillen:2006} define the mass ratio of the planet to the star ($\mu$) given the gap size to be $\mu = \left(\frac{\delta a}{1.5a}\right)^{7/2}$, where $a$ is the planet's semi-major axis and $\delta a$ is the difference between the planet's semi-major axis and the edge of the gap (i.e., half the gap width).  Given our best-fit gap parameters, the mass ratio $\mu = 0.00011\pm0.00004$.   Assuming the central star is an F4 spectral type with mass $\sim1.4$~$M_\odot$ yields an estimate for the planet mass of $0.16\pm0.06$~$M_\text{Jup}$.  \cite{Quillen:2006} estimate that for mass ratios of $10^{-4}$, $75\%$ of particles will be removed from the gap after $10^{3.4}$ orbits.  At a distance of 59~AU from the star, the orbital period is $\sim400$~years, which implies that the gap should be $>75\%$ depleted after $\sim1$~Myr.  Since the HD~15115 system is older ($\sim20-45$~Myr), it is plausible that a $0.16$~$M_\text{Jup}$ planet could produce a nearly depleted gap within the lifetime of the system. 

To demonstrate that a $\sim0.2$~$M_\text{Jup}$-mass planet could indeed open the observed gap in the HD~15115 disk, we performed N-body simulations using REBOUND \citep{Rein:2012}.  Figure~\ref{fig:fig4} shows the resulting face-on disk image and radial surface density profile after 1~Myr for a $0.2$~$M_\text{Jup}$-mass planet at semi-major axis 58~AU with eccentricity $e=0$ (top) and $e=0.06$ (bottom).  Both models assume a 1.4~$M_\odot$ star and include 10000 particles.  The planet quickly opens a gap with FWHM $\sim15$~AU in both models, but differences between the circular and eccentric cases could be revealed by future observations with increased sensitivity and resolution.  For a circular orbit, material is maintained in a co-rotating ring at the 1:1 resonance with the planet; these orbits are unstable in the eccentric case.  In addition, the gap appears slightly broader for an eccentric planet, indicating that a lower mass planet could maintain the observed gap.  Indeed, a lower mass planet could work in both cases, since \cite{Nesvold:2015} show that gaps grow over time due to collisional erosion.  Future high-quality extreme AO observations from a GPI-2.0 \citep{Chilcote:2018}, SCExAO \citep{Currie:2018}, or later ELTs could recover the gap in the HD~15115 disk and possibly detect a planet responsible for sculpting it.

It is well-known that young protoplanetary disks exhibit multiple rings, potentially created by forming planets \cite[i.e., the DSHARP survey, ][and references therein]{Andrews:2018}.  Since debris disks are the later evolutionary stage of circumstellar disks, we might expect them also to exhibit multiple rings.  However, to date, only two previous disks have shown evidence for multiple cold belts, HD~107146 \citep{Ricci:2015,Marino:2018} and HD~92945 \citep{Marino:2019}.  With the addition of HD~15115, we are beginning to build a sample of multiple ring debris disks that can inform our understanding of how and when disk gaps are produced.

\section{Conclusions}
\label{sec:conc}

We present new ALMA observations of the HD~15115 debris disk at 1.3 mm (230 GHz), which provide the highest resolution image of this unique system at millimeter wavelengths to date.  The ALMA image shows two bright peaks or ansae on either side of the star, characteristic of a double-ringed system viewed edge-on.  We fit models to the millimeter visibilities within a MCMC framework to robustly constrain the structure and geometry of the system.  The best-fit model has a gap at $58.9\pm4.5$~AU ($1\farcs2\pm0\farcs10$) with a width of $13.8\pm5.6$~AU ($0\farcs28\pm0\farcs11$) and a fractional depth of $0.88\pm0.10$.  There is no evidence from the ALMA data for the dramatic east-west asymmetry seen in scattered light images.  Since the mechanism producing this asymmetry appears to only affect small grains, we conclude that ram pressure stripping from an interaction with the local ISM is most likely.  From dynamical modeling, we conclude that the depleted gap in the disk is likely carved by a $0.16\pm0.06$~$M_\text{Jup}$ mass planet.  Higher resolution observations could reveal additional substructures in the disk resulting from interactions between the planet and disk material that would improve constraints on the planet mass and orbital properties.

\vspace{1cm}
We thank Christopher Stark for useful discussions of disk structure and sculpting mechanisms.  M.A.M. acknowledges support from a National Science Foundation Astronomy and Astrophysics Postdoctoral Fellowship under Award No. AST-1701406. A.M.H. acknowledges support from NSF grant AST-1412647. S.R. acknowledges support from NSF grant AST-1313268.  This paper makes use of the following ALMA data: ADS/JAO.ALMA \#2015.1.00633.S. ALMA is a partnership of ESO (representing its member states), NSF (USA) and NINS (Japan), together with NRC (Canada) and NSC and ASIAA (Taiwan) and KASI (Republic of Korea), in cooperation with the Republic of Chile. The Joint ALMA Observatory is operated by ESO, AUI/NRAO and NAOJ. The National Radio Astronomy Observatory is a facility of the National Science Foundation operated under cooperative agreement by Associated Universities, Inc.  This work has also made use of data from the European Space Agency (ESA) mission {\it Gaia} (\url{https://www.cosmos.esa.int/gaia}), processed by the {\it Gaia} Data Processing and Analysis Consortium (DPAC, \url{https://www.cosmos.esa.int/web/gaia/dpac/consortium}). Funding for the DPAC has been provided by national institutions, in particular the institutions participating in the {\it Gaia} Multilateral Agreement.  


\bibliography{References.bib}


\begin{deluxetable}{l l c c}
\tablecolumns{3}
\tabcolsep0.1in\footnotesize
\tabletypesize{\small}
\tablewidth{0pt}
\tablecaption{Best-fit Model Parameters \label{tab:results}}
\tablehead{
\colhead{Parameter} & 
\colhead{Description} & 
\colhead{Two Ring Model} &
\colhead{Gaussian Gap Model} 
}
\startdata
$F_\text{disk,1}$ & Total disk flux density [mJy] & $0.38\pm0.08$ & $-$ \\
$R_\text{in,1}$ & Ring 1 inner edge [AU] & $44.6\pm4.5$ ($0\farcs90\pm0\farcs09$) & $-$ \\
$R_\text{out,1}$ & Ring 1 outer edge [AU] & $50.9\pm8.8$ ($1\farcs04\pm0\farcs18$) & $-$ \\
$x_1$ & Ring 1 power law gradient & $-0.95\pm0.64$ & $-$ \\
$F_\text{disk,2}$ & Total disk flux density [mJy] & $1.61\pm0.09$ & $-$ \\
$R_\text{in,2}$ & Ring 2 inner edge [AU] & $65.7\pm4.5$ ($1\farcs34\pm0\farcs09$) & $-$ \\
$R_\text{out,2}$ & Ring 2 outer edge [AU] & $92.8\pm3.1$ ($1\farcs89\pm0\farcs06$) & $-$ \\
$x_2$ & Ring 2 power law gradient & $-0.65\pm0.78$ & $-$ \\
\hline
$F_\text{disk}$ & Total disk flux density [mJy] & $-$ & $1.98\pm0.03$ \\
$R_\text{in}$ & Ring inner edge [AU] & $-$ & $43.9\pm5.8$ ($0\farcs89\pm0\farcs12$) \\
$R_\text{out}$ & Ring outer edge [AU] & $-$ & $92.2\pm2.4$ ($1\farcs88\pm0\farcs49$) \\
$x$ & Ring power law gradient & $-$ & $-0.43\pm0.88$ \\
$R_\text{gap}$ & Gap location [AU] & $-$ & $58.9\pm4.5$ ($1\farcs20\pm0\farcs10$) \\
$W_\text{gap}$ & Gap FWHM [AU] & $-$ & $13.8\pm5.6$ ($0\farcs28\pm0\farcs11$) \\
$\Delta_\text{gap}$ & Gap fractional depth & $-$ & $0.88\pm0.10$ \\
\hline
$F_\text{pt}$ & Central point source flux density [mJy] & $0.04\pm0.01$ & $0.04\pm0.01$ \\
$\Delta\alpha$ & RA offset of star from disk centroid [$\arcsec$] & $0.10\pm0.05$ & $0.09\pm0.05$ \\
$\Delta\delta$ & DEC offset of star from disk centroid [$\arcsec$] & $-0.05\pm0.05$ & $-0.06\pm0.05$ \\
$i$ & Disk inclination [$\degr$] & $86.3\pm0.4$ & $86.2\pm0.5$\\
$PA$ & Disk position angle [$\degr$] & $278\pm1$ & $278\pm1$
\enddata
\end{deluxetable}

\end{document}